\documentclass[12pt]{spieman}  
\usepackage{amsmath}
\usepackage{graphicx}
\usepackage{setspace}
\usepackage{tocloft}
\usepackage{epstopdf}
\usepackage{cite}
\usepackage{rotating}
\usepackage{multirow}
\usepackage[caption=false]{subfig}
\usepackage[T1]{fontenc}
\usepackage{color}
\title{Scalable Platform for Adaptive optics Real-time Control (SPARC) Part 2: Field Programmable Gate Array (FPGA) implementation and performance}

\author[a,*]{Avinash Surendran}
\author[b]{Mahesh P. Burse}
\author[b]{A. N. Ramaprakash}
\author[a]{Padmakar S. Parihar}
\affil[a]{Indian Institute of Astrophysics, Koramangala 2nd Block, Bangalore - 560034, India}
\affil[b]{Inter-University Centre for Astronomy and Astrophysics, No. 4, Ganeshkhind, Pune University Campus, Pune - 411007, India}

\cftpagenumbersoff{figure}
\cftpagenumbersoff{table} 
\begin{document} 
\maketitle

\begin{abstract}
The next generation of Adaptive Optics (AO) systems on large telescopes will require immense computation performance and memory bandwidth, both of which are challenging with the technology available today. The objective of this work is to create a future-proof adaptive optics platform on an FPGA architecture, which scales with the number of subapertures, pixels per subaperture and external memory. We have created a scalable adaptive optics platform with an off-the-shelf FPGA development board, which provides an AO reconstruction time only limited by the external memory bandwidth. SPARC uses the same logic resources irrespective of the number of subapertures in the AO system. This paper is aimed at embedded developers who are interested in the FPGA design and the accompanying hardware interfaces. The central theme of this paper is to show how scalability is incorporated at different levels of the FPGA implementation. This work is a continuation of Part 1 of the paper which explains the concept, objectives, control scheme and method of validation used for testing the platform.
\end{abstract}

\keywords{Adaptive Optics, FPGA, Real time control}

{\noindent \footnotesize*}Avinash Surendran,  \linkable{asurendran89@gmail.com} 

\begin{spacing}{2}   

\section{Introduction}
\label{sect:intro}
Adaptive optics (AO) can correct for the aberrations in the optical wavefront due to turbulence in the atmosphere, and can help ground-based telescopes make observations with an angular resolution approaching that of a space-based telescope with a similar aperture size. Most of the next generation of AO systems on large telescopes are planned with pupil plane wavefront sensors (WFS) and actuator based deformable mirrors (DM). Matrix vector multiplication (MVM) is the preferred method for computing the DM shape from the WFS pixel inputs. The number of floating point operations (FLOPS) required for least square reconstructors and modal reconstructors {scales} with $n^2$, where $n$ represents the degrees of freedom of the AO system\cite{gavel2002}. The number of FLOPs can be brought down to $n \times log(n)$ for Fourier-based reconstructors with the disadvantages of a complex algorithm, special case requirements and a relatively higher error propagation\cite{poyneer2002}. The first-light laser guide star AO system on the Thirty Metre Telescope (TMT), called the Narrow Field Infrared Adaptive Optics System (NFIRAOS), will be required to solve a $35000\times7000$ MVM within 1 ms. This requires a memory bandwidth close to 800 GB/s and a computational power of 1.5 TeraFLOPS\cite{wang2013}. The computational resources necessary for a large-scale AO system still remain a non-trivial problem.

Early hardware designs\cite{hovey2010,boyer2011,zhang2012} incorporated FPGAs as possible candidates for implementing AO reconstruction on the NFIRAOS real-time controller (RTC). FPGAs and Graphical Processing Units (GPUs) have been competing from their inception to cater to the high performance computation market. GPUs excel in cost, floating-point processing and development costs while FPGAs are known for their predictable timing latency, low power consumption and high-speed interfacing options\cite{berten2016,falsafi2017}. A recent stagnation in compelling FPGA hardware, the high initial programming overhead for FPGAs and the rapid evolution of GPU and Central Processing Unit (CPU) performance has resulted in GPUs and CPUs dominating the choice of computing hardware for NFIRAOS\cite{wang2013,veran2014,smith2014} {and E-ELT\cite{bitenc2016,sevin2014} AO} over the years. The cost of hardware and development makes GPUs a compelling hardware choice for AO. The most recent hardware benchmarking results point to CPUs and Intel Xeon Phi coprocessors being preferred over GPUs\cite{smith2016,kerley2016}. The European Southern Telescope (ESO) has also moved on from a combination of FPGAs and Digital Signal Processors (DSPs) for its AO RTC\cite{fedrigo2006} to a heterogenous architecture involving CPUs and GPUs for mathematical processing and FPGAs for the high-speed interfacing\cite{dipper2013,basden2013}. {A similar scalable AO kernel involving an FPGA-based Network Interface Card (NIC) for peer-to-peer communication, with GPUs performing the intensive computational part, is explored in Perret et al\cite{perret2016}. A more recent approach from ESO (called Green Flash) aims to compare competing technologies of co-processors, GPUs and FPGAs for use as the computing platform for the AO system on the E-ELT\cite{gratadour2016}. }

With the advent of serial memory like the Hybrid Memory Cube (HMC)\cite{pawlowski2011} and Intel's acquisition of Altera, there has been a resurgence of interest in the use of FPGAs as hardware accelerators. e.g. the Intel Stratix 10 is expected to have 5,000 dedicated floating point units, 1 TB/s of external serial memory bandwidth and a 32-bit peak floating point throughput of 9.2 TeraFLOPS\cite{nurvitadhi2017}. This performance is comparable to the Nvidia Titan X Pascal GPU, which is one of the most powerful GPUs in the market today. Other recent innovations by Xilinx\cite{xilinx2016,xilinx2017} also make FPGAs competitive with GPUs and CPUs in terms of performance. In light of new developments, E-ELT is considering the use of FPGAs for its AO RTC\cite{felini2016}. Rapid deployment of AO on a network of small telescopes is also possible with an AO RTC implemented on an FPGA platform\cite{cegarra2016}, {as it helps in the reduction of programming overheads by acting as a plug-and-play AO-on-a-chip.}

We have developed an architecture which overcomes the primary weakness in FPGAs namely, the special skills programming need, with the implementation of this platform. SPARC is designed to form the basis to create a standard FPGA code framework for AO in the future. The standard framework with the option of modular changes in the core algorithm reduces the programming overhead. Compatibility across FPGA families is ensured by programming in the native VHDL language, while resorting to hardware specific blocks only for a small fraction (25\%) of the FPGA resources. FPGA based hardware architectures for wavefront reconstruction have been attempted\cite{kepa2008,lynch2008}, but there have been few efforts to create a truly scalable platform\cite{goodsell2005,goodsell2006,valles2012,gratadour2016}, that also offers flexible interfaces for the input pixel stream from the WFS as well as output control stream for the DM. {Part 1\cite{SPARC1} of this paper explains the primary objective of the work, the computational hardware required for an AO system on a large telescope, architecture, methods of validation (including the interface with a real AO system) and the results from the same. Part 1\cite{SPARC1} is targeted at astronomers and the users who are interested in understanding the AO performance of SPARC. Part 2 (this paper) describes in detail, the implementation of SPARC on an off-the-shelf FPGA development board. This part is targeted at FPGA developers and electronics engineers.}

Section~\ref{sec:assum} describes the assumptions that we made in designing the platform. Section~\ref{sec:resan} explains the analysis of the minimum binary {precision} required for representing the reconstruction matrix while keeping the computational errors within acceptable limits. Section~\ref{sec:impl} gives the complete implementation map of the AO kernel. It includes the dataflow, necessary timing synchronizations and the core reconstruction algorithm with the levels of scalability incorporated within the same. Section~\ref{sec:res} consists of the results pertaining to the AO reconstruction time and the logic resources used by the FPGA. Section~\ref{sec:disc} explores future work and a possible roadmap for AO on FPGAs.

\section{Assumptions}
\label{sec:assum}
The assumptions have been derived from the objectives mentioned in Part 1\cite{SPARC1} of this work. We have assumed that the reconstruction matrix maps from a pupil plane wavefront sensor (WFS) to an actuator based deformable mirror (DM), as is the case for the majority of AO systems deployed around the world. A Shack-Hartmann Wavefront Sensor (SHWFS) model is used with a Centre of Gravity (CoG) algorithm for slope computation. {This CoG algorithm is modular and implemented as a stand-alone function executed in a single clock cycle in the program, and can be changed to any mathematical operation on a square array of pixels. This makes the algorithm adaptible with little programming effort to say, a pyramid WFS where a similar mathematical operation is executed for a square array of pixels (taken in a different format from the SHWFS).} A strict implementation of Fried geometry and a conventional MVM with an ideal reconstruction matrix is considered for testing the platform. 

As the reconstruction matrix is fed from an external source and generated from outside the FPGA, other geometries can be easily implemented. As the conversion of phase outputs to DM controller input is not resource-intensive and changes from controller to controller, it is assumed to be done outside of the FPGA. The pixels are considered to be coming through a single channel with the camera controller rearranging the pixels to be sent in the correct order. Again, this can be easily adapted to cater to other streaming scenarios of pixel data.

\section{{Precision} Analysis}
\label{sec:resan}
To reduce the complexity of implementation and to avoid the usage of device specific floating- point IP cores, we use fixed-point arithmetic for slope computation and AO reconstruction. The number of bits with which each variable is represented decides the rounding-off error which propagates through the computation process. This final error can become substantial considering the large number of multiply-and-accumulate (MAC) operations in AO reconstruction. 

\begin{figure}
\centering
\includegraphics[width=0.9\textwidth]{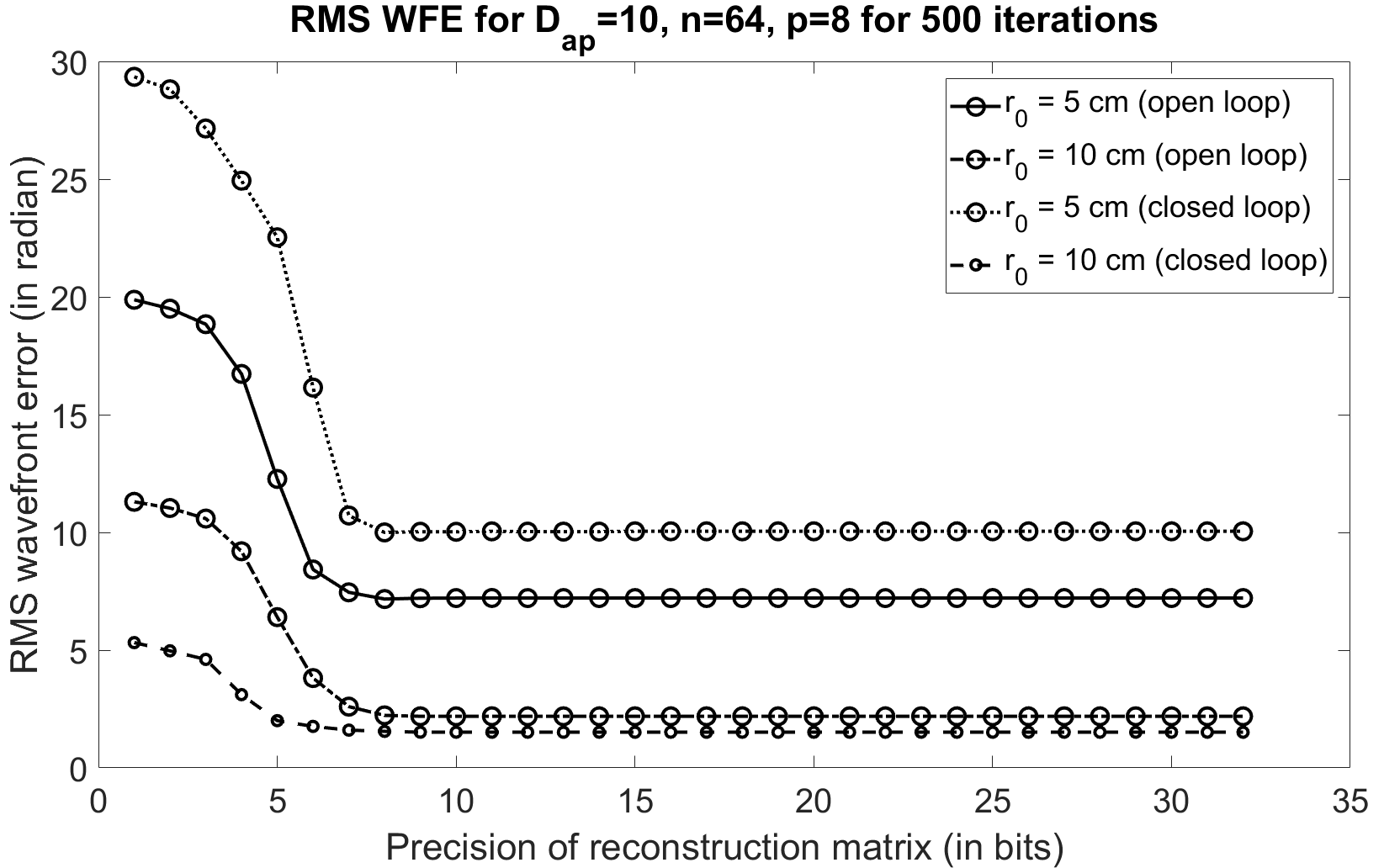}
\caption{Comparison between the RMS error between the input and the reconstructed WF, and the number of bits used after the decimal point to represent the reconstruction matrix.}
\label{fig:ra}
\end{figure}

We conducted an analysis of the {precision} in number of bits that would be required to represent the reconstruction matrix without significant error in the phase outputs. As the other variables (phase, slope etc.) do not need a large amount of memory, they are represented by 32 bits and are stored in the internal memory of the FPGA. A Von-Karmann phase spectrum (modeled on the one created by Sedmak\cite{sedmak2004}) was used as the input for modeling the atmospheric turbulence. We modeled the error propagation due to AO reconstruction for different values of the primary aperture diameter, number of subapertures and the Fried parameter. Fig.\ref{fig:ra} shows the plot of the RMS residual error of the difference between the input and the reconstructed wavefront (WF), against the number of bits used after the decimal point to represent the reconstruction matrix. A phase screen of size 10 m with {seeing conditions corresponding to a Fried parameter of 5 cm and 10 cm (open and closed loop)} are used for simulating an AO system having $64\times64$ subapertures with $8\times8$ pixels per subaperture. 

The RMS WF error remains constant beyond 8 bits of {precision}. As errors in WF sampling and slope computation could propagate rapidly through the reconstruction matrix for bit accuracies lower than 8, each reconstruction matrix element was chosen to be stored as a 16-bit fixed point value.

\section{Implementation}
\label{sec:impl}

The platform requires an FPGA chip for the functioning of the AO loop and a DDR-type memory for storing the reconstruction matrix. The interface with a WFS or DM varies widely with the type of hardware used, hence we have created a flexible wrapper to accommodate different interfaces. The implementation of the AO kernel is divided into three sub-sections. The arrows in the Dataflow sub-section show the connectivity between the different functional blocks.  The timing synchronization explains how the platform adapts to the two different timing scenarios which arise as we scale up the size of the AO system. The third sub-section explains how the platform is made scalable to enable the interfacing of DDR memories of varying clock frequencies and data widths. It also explains the constraints imposed by the type of memory (which is used to store the reconstruction matrix) and the logic resources in the FPGA.

\begin{sidewaysfigure}
\includegraphics[width=1\textwidth]{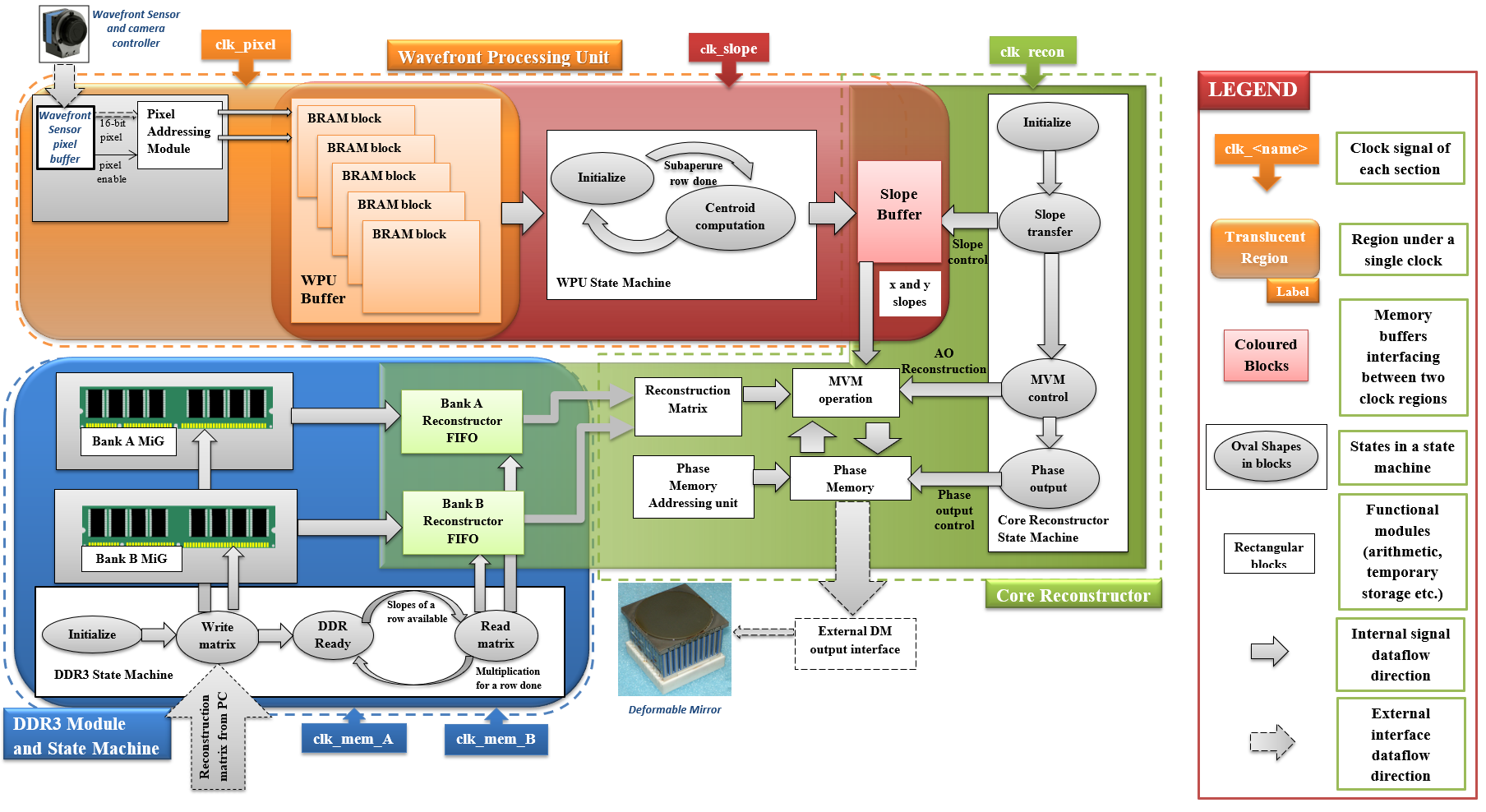}
\caption{Dataflow with hardware interconnections of the platform. The figure shows three state machines: {\lq{Wavefront Processing Unit}\rq} acquires the pixels and computes the slopes; the {\lq{Core Reconstructor}\rq} controls the other state machines and performs AO reconstruction; and the {\lq{DDR3 state machine}\rq} creates a scalable pathway between the DDR3 memory and the AO Reconstructor for the reconstruction matrix. The colour with which each section is shaded varies according to the clock which drives the particular section.}
\label{fig:df}
\end{sidewaysfigure}

\subsection{Dataflow}
\label{subsec:df}
Fig.~\ref{fig:df} shows the broad division of the platform into three state machines and five clock regions, each of which are shaded by a different colour. The clock regions only overlap at buffer regions, where data is input to the buffer at one clock frequency and output at another clock frequency. The three state machines directly map to the state machine implementation described in Part 1\cite{SPARC1} of this paper. The state machines are driven by different clock frequencies, according to the function and external interface to which the state machine exchanges the data with.

\subsubsection{Pixel clock ($clk\_pixel$)}
The Wavefront Processing Unit (WPU) state machine is responsible for simultaneously reading in the pixels from the WFS and the pipelined computation of slopes. The pixels from the WFS CCD are read into a temporary buffer, and the interface is isolated by the WPU. This is required as the interfaces of the different WFS cameras vary widely and the external interface needs to be isolated from the computational section of the WPU, both of which may work at different clock frequencies. We could choose a clock frequency of 100 MHz for the pixel clock because it did not drive any state machines or complex arithmetic. The number of slopes to be computed per clock cycle (of the slope clock) is set using a parameter named {\lq{iter}\rq}. The {\lq{WPU Buffer}\rq} (shown in Fig.~\ref{fig:df}) is created by the WPU to store the pixels in an arrangement which will facilitate the parallel readout of the pixel values for slope computation (as explained in Section~\ref{subsubsec:clk_slope}).

\subsubsection{Slope clock ($clk\_slope$)}
\label{subsubsec:clk_slope}
The WPU state machine enters the {\lq{centroid computation}\rq} mode after the pixels, corresponding to a single row of subapertures, are transferred to the WPU buffer. The number of slopes computed in a single clock cycle is dependent on the nature of the FPGA and can be changed through the variable {\lq{iter}\rq}. The arrangement of the pixels in the WPU buffers allow for this capability. For e.g. a sub-\$100 Xilinx Spartan 6 will be able to compute 2-8 slopes/clock cycle while a Xilinx Virtex-7 chip which costs more than \$5000 will able to compute upto 64 slopes/clock cycle. Any division operation (like the one in the CoG algorithm) consists of cascaded subtractions and is difficult to be parallelized, leading to a maximum clock frequency of only 12.5 MHz for the slope clock. The parallel computation of slopes for several subapertures (within a row) compensates for the reduced clock frequency to a certain level. The time taken for slope computation is a fraction of the WFS frame time and the time needed for AO reconstruction, and hence does not affect the overall latency of the AO loop.

\subsubsection{Reconstructor clock ($clk\_recon$)}
The core reconstructor state machine communicates with the WPU state machine to start pixel acquisition when the reconstruction matrix is written into the DDR3 memory and all other subsystems become ready. The core Reconstructor state machine enters the {\lq{slope transfer}\rq} state when the slopes corresponding to the first row of subapertures become ready. It instructs the DDR3 state machine to start reading out the entries from the relevant section of the reconstruction matrix, to be multiplied with the available slopes. The {\lq{MVM operation}\rq} block accepts the temporary phase values, the x-slopes and y-slopes of a particular subaperture, and the corresponding elements of the reconstruction matrix to complete a single multiply-and-accumulate (MAC) operation. The new phase values computed from the MAC operation are stored in the phase memory. The process is continued till all the slopes are multiplied with the elements of the complete reconstruction matrix. The number of multiplications in a single clock cycle is determined by the number of reconstruction matrix elements which are extracted from the DDR3 memory, and the number of Digital Signal Processing (DSP) units available. DSP units are fast specialized MAC hardware present inside the FPGA, and can perform the MAC operation more efficiently and faster compared to conventional logic inside the FPGA. The MVM equation and the implementation of scalability from the DDR3 memory to the MVM operation is mentioned in Section~\ref{subsec:recon}. The reconstructor clock is set to a clock frequency of 50 MHz considering the time required for a full MAC operation in the DSP.

\subsubsection{Memory clock ($clk\_mem$)}
Two banks of DDR3 memory are used to store the two halves of the reconstruction matrix corresponding to the x-slopes and y-slopes respectively. Unlike internal Block RAM, there is an unpredictable delay between data request and availability for DDR3 memory. The function of the DDR3 state machine is to isolate the random delay between corresponding reconstruction matrix elements coming from the two DDR3 memory banks, from the rest of the platform. The DDR3 state machine also ensures that it can interface different DDR3 memory hardware with different bandwidths and data widths. E.g. The current Memory Interface Generator (MiG) provided by Xilinx for DDR3 memory on the VC-709 board works at a clock frequency of 200 MHz with a data width of 512 bits. The DDR3 state machine converts this data stream into a data width of 2048 bits at a clock frequency of 50 MHz to interface with the core reconstructor.

\subsection{Timing synchronization \label{subsec:ts}}
\begin{figure}
\centering
\includegraphics[width=1\textwidth]{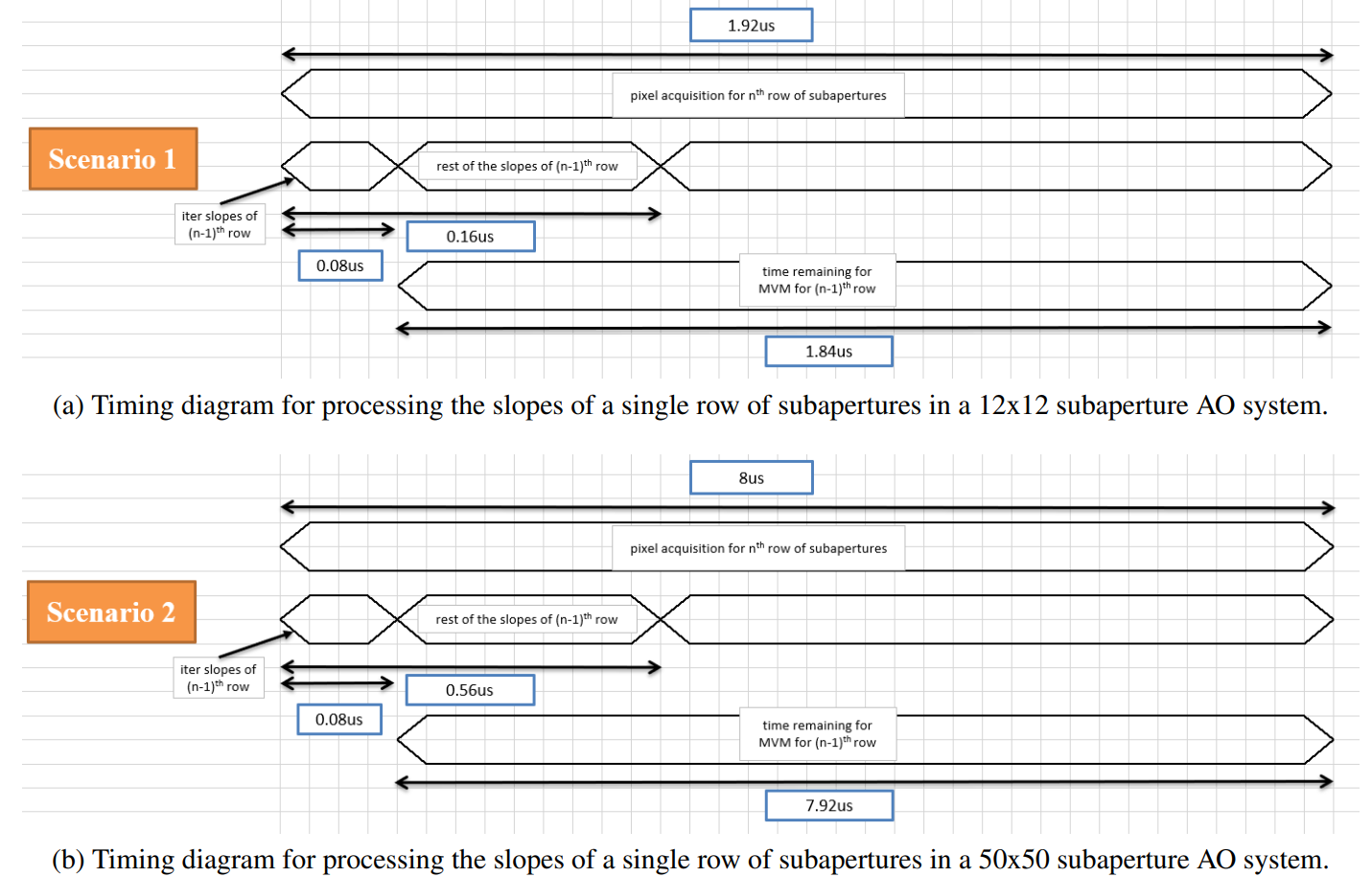}
\caption{{{\lq{iter slopes of $(n-1)^{th}$ row}\rq} refers to the first clock cycle of slope computation, after the acquisition of pixels corresponding to a single row of subapertures. The number of slopes to be computed in a clock cycle (and hence the time taken to compute the slopes) can be set by the user according to the performance of the FPGA used. When the processing of {\lq{iter}\rq} slopes gets completed, AO reconstruction can start for the available slopes while the {\lq{rest of the slopes of $(n-1)^{th}$ row}\rq} are computed. The {\lq{time remaining for MVM}\rq} time shown is the remaining time available to complete AO reconstruction before the next row ($n^{th}$) of subaperture pixels are ready to be processed.}}
\label{fig:scen}
\end{figure}

If we consider Fried geometry with $n \times n$ subapertures, the number of reconstruction matrix elements which are required to be multiplied with a single slope is $(n+1)^2$. For a single row of subapertures in a small AO system, the time elapsed for retrieving the reconstruction matrix elements is less than the combined time elapsed for reading out the pixels and computing the slopes of the next row of subapertures. e.g. Fig~\ref{fig:scen}a shows the timing diagram of AO reconstruction for a single row of subapertures in a $12\times12$ subaperture system. {The pixel readout for the $n^{th}$ row of subapertures is shown in Fig~\ref{fig:scen}a, with the slope computation and the MVM operation of the $(n-1)^{th}$ row of subapertures (after the pixels corresponding to the $(n-1)^{th}$ row of subapertures have been read by the FPGA).} We are assuming $4\times4$ pixels per subaperture with each pixel being read out of a buffer every 10 ns. In the current example, eight slopes are generated at every clock pulse of 80 ns. The time remaining for multiplying the slopes of the row of subapertures with the corresponding reconstruction matrix elements is 1.84 $\mu$s. For that single row of subapertures, the actual time taken for the readout of the elements from a 64-bit DDR3 RAM operating at 1600 MHz and the corresponding multiplication is only 160 ns. Hence, the entire AO reconstruction gets completed almost immediately after the last row of subapertures is read in.

For a $50\times50$ subaperture system with the same number of pixels per subaperture, clock signals and development board, AO reconstruction for a row of subapertures take 24-26 $\mu$s. In this case (Fig~\ref{fig:scen}b), the entire AO reconstruction will take longer than that is required for reading out all the pixels from the WFS. The top module makes the platform scalable by adapting to the two scenarios to ensure the correct functioning of the AO reconstruction pipeline.

\begin{sidewaysfigure}
\includegraphics[width=1\textwidth]{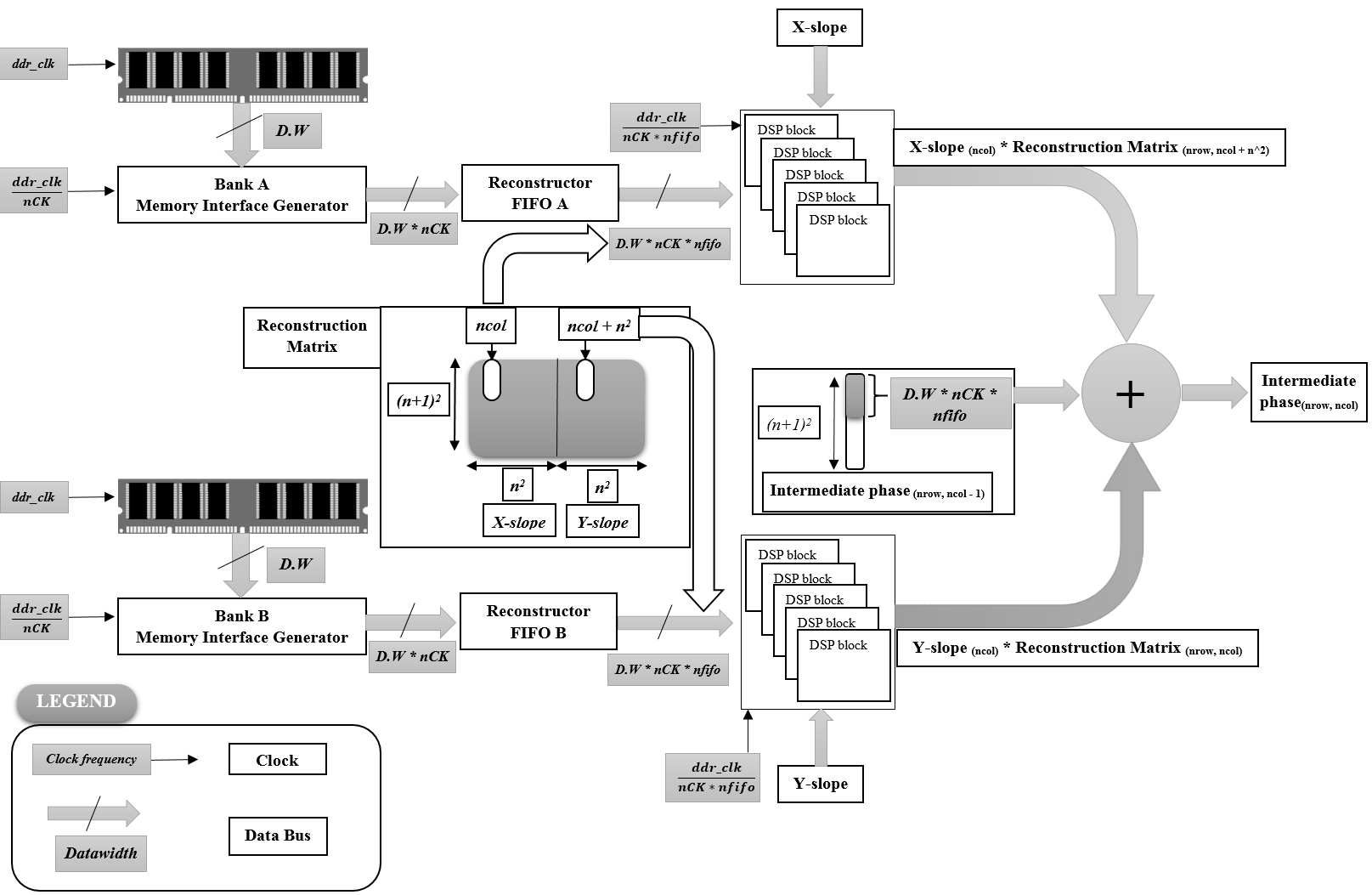}
\caption{Reconstruction data pathway showing the dataflow from the DDR3 memory modules to the core reconstruction algorithm
\label{fig:recon_core}}
\end{sidewaysfigure}

\subsection{Scalability of the reconstruction algorithm} 
\label{subsec:recon}

Fig.~\ref{fig:recon_core} provides a detailed schematic of the flow of data from the DDR3 memory to the MVM operation and explains how scalability is implemented in the FPGA logic. The reconstruction matrix is split into two parts to be multiplied with the x-slopes and the y-slopes. Bank A stores the x-slope section and Bank B stores the y-slope section of the reconstruction matrix. The total size of the matrix is $(n+1)^2 \times 2n^2$. Although we have used DDR3 memory, the FPGA logic is compatible with any DDR memory coupled with a compatible MiG provided by Xilinx. This would be useful for implementing a small AO system using cheaper FPGAs which are compatible with DDR and DDR2 memory. We assume a generalized DDR clock frequency of $ddr\_clk$. The Xilinx MiG converts the DDR clock frequency of $ddr\_clk$ (1600 MHz in our case) to $\frac{ddr_clk}{nCK}$ (200 MHz in our case), while simultaneously increasing the data width by $nCK$ times (8 in our case) to preserve bandwidth. The frequency conversion is determined by Xilinx and can vary with the nature of memory used and the FPGA. The role of the reconstructor FIFO is to output the reconstruction matrix elements at the clock frequency of the core reconstruction algorithm. For the VC-709 board and a non-pipelined DSP architecture, we can perform the AO reconstruction at 50 MHz. If the frequency at which the AO reconstruction is done is denoted by $recon\_clk$,
\begin{align*}
recon\_clk &= \frac{ddr\_clk}{nCK \times nfifo} \\
nfifo &=\frac{ddr\_clk}{nCK \times recon\_clk}
\end{align*}
where $nfifo$ is a constant in the program which can be varied to configure the FIFO to work seamlessly with any DDR IP core and with any FPGA. For the VC-709 development board, the DDR3 IP core frequency ($\frac{ddr\_clk}{nCK}$) is 200 MHz, and the AO reconstruction happens at 50 MHz ($recon\_clk$), which requires $nfifo$ to be set to 4.

The same dataflow ensures that the reconstruction matrix elements are continuously being read out from the DDR memory even though the AO reconstruction works at a lower frequency. If we assume that the reconstruction matrix elements (or data words) read out from the DDR RAM every clock cycle is $D.W.$, the number of DSPs instantiated is $2 \times D.W \times nCK \times nfifo$. For our prototype on the VC-709 board, 256 parallel DSPs are used every clock cycle. At every clock cycle, the core reconstructor instantiates $D.W \times nCK \times nfifo$ parallel instances of the following operation (performed by the parallel array of DSP blocks shown in Fig.~\ref{fig:recon_core}):
\begin{multline*}
Phase(nrow,ncol) = [recon\_mat(nrow,ncol) \times x\_slope(ncol)] \\
+ [recon\_mat(nrow,ncol+n^2) \times y\_slope(ncol)] + Phase(nrow,ncol - 1)
\end{multline*} 
where $recon\_mat$ is the reconstruction matrix stored as a vector in DDR memory, $x\_slope$ is the slope value of each subaperture in the x-direction, $y\_slope$ is the slope value of each subaperture in the y-direction, $nrow$ is the row number of the reconstruction matrix element and $ncol$ is the column number of the reconstruction matrix element. The operation continues till the value of $ncol$ reaches $n^2$. The {\lq{Phase}\rq} block is the corresponding intermediate phase value stored in the internal Block RAM of the FPGA after every MAC operation, as part of the bigger MVM operation.

The performance of the platform is determined by the performance of the DDR memory ($ddr\_clk$ and data width) and availability of logic in the FPGA (determined by the number of DSPs which are available in the FPGA).

\section{Results}
\label{sec:res}

{As explained in Section~\ref{sec:assum}, the system was validated through a hardware-in-the-loop (HIL) simulation over the Peripheral Connect Interface express (PCIe) interface. HIL simulation creates a range of atmospheric turbulence characteristics for different telescope aperture sizes and the performance of SPARC is tested for a range of subaperture sizes. This simulation measures the AO reconstruction time and verifies the phase outputs generated by SPARC. The HIL simulation could not provide a real-time simulation because of the large pixel processing time at the end of the host PC. Hence, an interface was built between the FPGA platform to connect with the WFS and the DM of the iRobo-AO system. The iRobo-AO system is closely modelled on Robo-AO\cite{baranec2012}, which was built throufh a collaboration between the Inter-University Centre for Astronomy and Astrophysics (IUCAA) and California Institute of Technology, USA. The iRobo-AO system is currently residing as a laboratory platform at IUCAA awaiting on-sky installation. More details about the HIL and iRobo-AO interface, along with results from the validation are provided in Part 1\cite{SPARC1} of the paper.}

SPARC is designed to be scalable with respect to the following parameters:
\begin{enumerate}
\item {Number of pixels and pixels per subaperture: The number of subapertures and pixels per subaperture are variables at the top level of the FPGA code, and any change in the variables will automatically cascade seamlessly to change the FPGA design (memory allocation, state machine behavior, DSP usage, reconfigurable IPs etc.) without any need to change the main program.}
\item {Memory bandwidth: As explained in Section~\ref{subsec:recon}, the external DDR memory is connected to a reconfigurable FIFO in the FPGA, which can adapt seamlessly to connect to different memory datawidths and frequencies.}
\item {Portability across FPGA families: The main program is developed using native VHDL (with no external IPs), which makes SPARC adaptable with the FPGAs produced by most companies around the world. The non-portable part constitutes about 25\%, which mainly involves the communication through external interfaces (DDR, PCIe etc.).}
\end{enumerate}

The results related to the loop time and the logic resources used by the platform are presented here. The results pertaining to AO time-series results are given in Part 1\cite{SPARC1} of the paper. Even for the AO system with the largest number of subapertures ($50\times50$) that we tested the platform with, the logic resources on the off-the-shelf inexpensive development board was not a constraint. 

\begin{figure}
\centering
\includegraphics[width=0.9\textwidth]{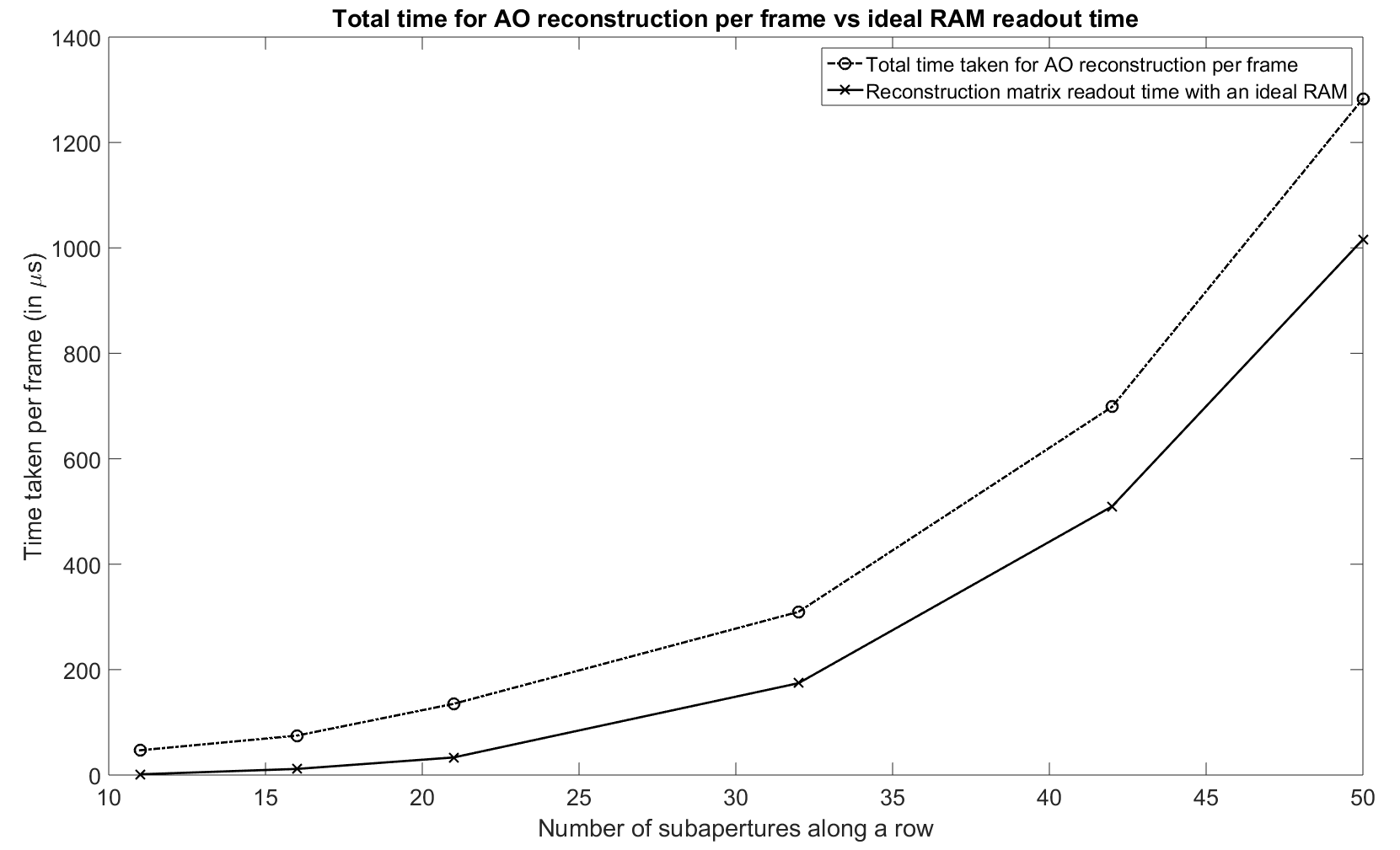}
\caption{Comparison of total time for AO reconstruction per frame vs speed of an ideal RAM (for different subaperture sizes)}
\label{fig:result_speed}
\end{figure}

Fig.~\ref{fig:result_speed} shows the comparison between the actual time taken for AO reconstruction per frame at different SHWFS subaperture sizes, with the time taken by an ideal DDR3 RAM (without delays or latencies) on the development board. The time taken by an ideal DDR3 RAM is given by the ratio of the size of the reconstruction matrix to the rated memory bandwidth. Although the state machine and the memory interface introduces latencies and delays, the performance is limited by the memory bandwidth. We have made the system scalable with the clock speed and datawidth of the DDR memory to make the platform compatible with faster memory modules. On the VC-709 development board, the complete AO reconstruction for a $50\times50$ subaperture frame (with $4\times4$ pixels per subaperture) takes 1.283 ms out of which about 1 ms is taken for fetching the reconstruction matrix from the DDR3 memory. If the DDR3 memory is replaced with a DDR4 memory with a bandwidth of 2400 MHz and a data width of 72 bits (similar to what is provided on the Bittware XUSP3S development board\cite{bittware2017}), the closed loop time for the AO system will reduce to around 950 - 1000 $\mu$s.

\begin{figure}
\centering
\includegraphics[width=0.9\textwidth]{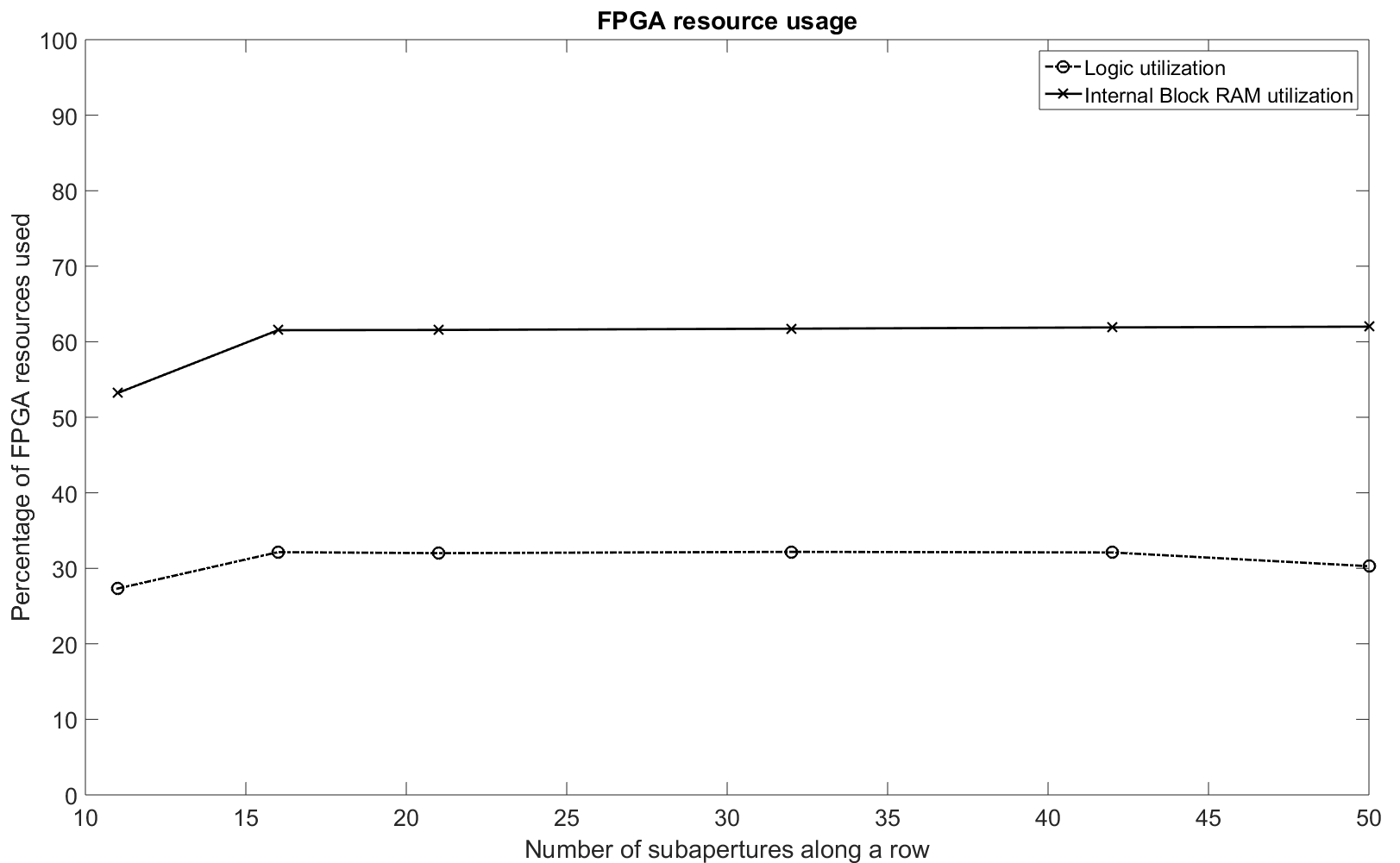}
\caption{Logic and Block RAM utilization by the FPGA (for different subaperture sizes)}
\label{fig:result_resources}
\end{figure}

Fig.~\ref{fig:result_resources} shows that the logic resource usage and internal Block RAM utilization stays constant for different values of subaperture sizes. The logic resource utilization is only dependent on the number of slopes computed per clock cycle\cite{asurendran2015} and the nature (speed and datawidth) of the DDR3 memory used. The configuration used for generating Fig.~\ref{fig:result_speed} and Fig.~\ref{fig:result_resources} computes 16 slopes per clock cycle. The platform uses a little more than 30\% of logic resources for the current configuration and meets the timing requirements of the FPGA on the development board.

\section{Discussion}
\label{sec:disc}
The competition between GPUs and FPGAs in the different fields of High Performance Computing platforms are intense, and the time lag between innovation to commercialization is quite small. Any software implementation on highly parallel platforms should not only be backward compatible but also be future proof and flexible to incorporate hardware which is yet to be envisioned. The work towards a scalable platform, which is FPGA agnostic and at the same time be flexible to incorporate future memory modules, was created with the vision to be flexible with regards to how fast the technology moves in the field.

We believe that the advent of serial memory modules like the HMC and the High Bandwidth Memory (HBM) will remove the memory bandwidth bottlenecks which limit the computational requirements of current large AO systems. Both HMC and HBM are the result of a combination of a three-dimensionally stacked memory with a high-speed serial interface (similar to multi-gigabit optical transceivers). {HMCs from Micron, which are commercially available from the beginning of 2017}, deliver a theoretical memory bandwidth of 160 GB/s per chip, which would be about 6-8 times the bandwidth of the fastest DDR-type memories available today. {HMCs are also available on FPGA development boards from at least three certified partners of both Xilinx and Altera. HBMs are currently available on high performance GPUs\cite{tech2017}, but are expected to be available on the next generation of Xilinx Ultrascale+ FPGAs\cite{xilinxhbm2017}.} Our future goal is to adapt this platform to a combination of Xilinx Ultrascale+\cite{xilinx2016,xilinx2017} FPGAs interfaced with a serial memory (when they become mainstream), which can enable the effortless implementation of Extreme-AO requirements for large telescopes. The attractiveness of the SPARC platform is that it can equally easily be used on less powerful FPGAs and memory modules which are slower and has a lower memory bandwidth, to cater to the needs of small telescopes.

\appendix    

\acknowledgments 
We are thankful to the Department of Science and Technology, University Grants Commission and the Infosys Foundation for funding and supporting this project.

\bibliography{SPARC_2}   
\bibliographystyle{spiejour}   


\listoffigures
\listoftables

\end{spacing}
\end{document}